\documentclass[a4paper,11pt]{article}

\usepackage[utf8]{inputenc}
\usepackage[T1]{fontenc}
\usepackage{amsmath,amssymb,amsthm}
\usepackage{graphicx}
\usepackage{booktabs,tabularx}
\usepackage[margin=1in]{geometry}
\usepackage[numbers,sort&compress]{natbib}
\usepackage{xurl}
\usepackage[hidelinks]{hyperref}
\usepackage[capitalise]{cleveref}

\title{Integrity and Credibility in Navigation:
From Error Characterization to Operational Assurance}
\author{Penggao Yan, Baoshan Song, Yuan Li, Ronghe Jin, Li-Ta Hsu\thanks{Corresponding author: Li-Ta Hsu (\href{mailto:lt.hsu@polyu.edu.hk}{\nolinkurl{lt.hsu@polyu.edu.hk}}).}}
\date{}
\hypersetup{%
  pdftitle={Integrity and Credibility in Navigation: From Error Characterization to Operational Assurance},
  pdfauthor={Penggao Yan, Baoshan Song, Yuan Li, Ronghe Jin, Li-Ta Hsu},
  pdfsubject={Technical perspective; working draft}
}

\begin{document}
\maketitle

\begin{abstract}
Navigation integrity and estimator credibility use error and uncertainty information to address different questions. This perspective distinguishes actual positioning error from reported statistical characterizations, formulates integrity as control of hazardous unwarned-use risk under the requirement of a specified operation, and describes credibility as the justification of an uncertainty or risk claim by its evidence usage and modeling process. Three hypothetical scenarios show how deteriorated accuracy, a contradicted uncertainty report, and a hazardous event can lead to different credibility and integrity judgments. The resulting assessment asks what happened, why the reported claim should be believed, and whether the information is sufficient for the intended operation.
\end{abstract}

\section{Introduction}
\label{sec:introduction}

Modern navigation systems estimate a user's position or state from available observations. The estimate itself, however, does not describe the complete navigation outcome: the difference between the estimate and the true state determines how far the reported position is from reality. In many applications, this result is used immediately for control, guidance, or an operating decision rather than evaluated only after the fact~\cite{reid2019localization}. Because the true state is usually unavailable during operation, the system cannot directly observe its current positioning error. The practical question is therefore more demanding than ``what position has been estimated?''; it is whether the navigation information remains suitable for the decision that depends on it. This question is especially consequential in safety-critical navigation, where continued use of information that no longer meets the task requirement can affect operational safety \cite{wgc2015milestone2}.

Precision approach in civil aviation provides a clear example. The operation depends continuously on navigation information whose performance must satisfy specified requirements \cite{gmv2011integrity,wgc2015milestone2}. Those requirements concern more than a small average position error: they include an allowable error tolerance, a limit on the probability of hazardous misleading use, and a deadline for warning the user when the information should no longer be used~\cite{esaNavigationRequirements}. The operational risk is consequently the risk that navigation information remains in use while a dangerous error is present and no effective warning arrives in time \cite{wgc2015milestone2,blanch2015baseline}. Navigation integrity addresses this question by relating the navigation service and its warning mechanisms to the risk requirement of the specified operation \cite{blanch2015baseline,pervan1998multiple,milner2020time}. A system with small current or average error cannot, by that fact alone, be declared to satisfy an integrity requirement.

Integrity monitoring cannot normally observe the true positioning error directly. It must instead use available statistics and models, such as reported covariance, residual information, assumed error distributions, dependence between error sources, fault hypotheses, and fault probabilities, to infer or bound the risk of a hazardous event~\cite{blanch2015baseline}. The resulting risk estimate, protection bound, or availability decision is therefore conditional on how well those descriptions represent the error-generating and fault processes encountered by the operation. If a variance is understated, an important correlation is omitted, a relevant fault mode is absent, or a tail assumption is inappropriate, an integrity calculation may be executed correctly while its connection to the real operation remains weak. Before an integrity result is used operationally, one must therefore ask why the error and uncertainty descriptions supporting it should be believed.

This evidence question is naturally discussed through credibility. In estimation, credibility concerns whether an estimator's self-assessed error statistics reliably describe its actual error statistics~\cite{li2002credibility,li2011evaluation}. In this paper, we extend this perspective to uncertainty and navigation-risk reports, considering both their statistical support and the assumptions and validation evidence on which they depend. Credibility and integrity are closely related because the latter may rely on the former's subject matter, but they are not the same assessment. Credibility concerns the justification of a report and its statistical meaning, whereas integrity concerns the risk of hazardous unwarned navigation use for a specified operation.

Both concepts have established research traditions~\cite{li2002credibility, li2011evaluation, milner2020time, blanch2015baseline,pervan1998multiple}, motivating an explicit comparison of their roles in navigation assessment. The central difficulty is to keep separate the object being evaluated, the probability statement being made, the evidence that can support that statement, and the way the statement enters an operational decision. An uncertainty interval with supported nominal coverage is not automatically a protection bound for a particular integrity risk requirement; conversely, one large error or one hazardous realization does not by itself establish that a credibility claim has failed. These distinctions matter because they change how accuracy, uncertainty reports, monitoring behavior, warnings, and continued use should be interpreted.

This paper clarifies the different roles and relationship of navigation integrity and estimator credibility. Its main contribution is a systematic comparison of what each concept evaluates, which probability and evidence claims it depends on, and why one cannot be used as a substitute for the other. Section~\ref{sec:error-reports} first distinguishes actual positioning error from reported error and uncertainty characterizations. Section~\ref{sec:integrity} then formulates integrity as an operation-dependent risk requirement and states the assumptions on which an integrity argument rests. Section~\ref{sec:credibility} discusses credibility, including evidence usage, model adequacy, and representative diagnostics. Section~\ref{sec:comparison} compares the two concepts through direct synthesis and three hypothetical scenarios, and Section~\ref{sec:conclusion} closes with the implications for navigation assessment.

\section{Positioning Error and Uncertainty Characterization}
\label{sec:error-reports}

\subsection{Position Estimate and Actual Error}

At time $t$, let $x_t$ denote the true position component and let $\hat{x}_t$ be the value estimated from the available observations. The realized positioning error is
\begin{equation}
\label{eq:error}
e_t=x_t-\hat{x}_t.
\end{equation}
This quantity describes what actually happened at that time. It is determined jointly by the true state and the estimate, and it is therefore different from an uncertainty measure produced by the estimator.

When a sufficiently accurate reference is available in an experiment or a retrospective analysis, $e_t$ can be evaluated and the realized positioning performance can be described. During normal operation, however, the true state $x_t$ is usually not directly available. The current value of $e_t$ is consequently not an online observable on which the navigation system can base its use decision.

\subsection{Reported Error and Uncertainty Information}

Because the realized error is unavailable online, an estimator normally reports information intended to characterize its error. A report may contain a mean or bias estimate, a standard deviation or covariance, a probability region, or a bound indexed by a nominal probability level. For example, fixed reported values $\mu_t^{\mathrm{rep}}$ and $s_t\geq0$ can state the intended first- and second-moment characterization
\begin{equation}
\label{eq:moments}
\mathbb{E}[e_t]=\mu_t^{\mathrm{rep}},\qquad \operatorname{Var}(e_t)=s_t^2.
\end{equation}
These are statements about the error distribution; neither $\mu_t^{\mathrm{rep}}$ nor $s_t$ is the realized error itself. A different report may specify a radius $r_t(\alpha)$ and claim that the error is contained with a stated probability, for example $\Pr(|e_t|\leq r_t(\alpha))\geq1-\alpha$. 

Since the current $e_t$ cannot generally be observed, a navigation system must use reported uncertainty, statistical error models, and other monitoring information when it judges whether a result remains reliable enough for a task. In an ordinary application this may support an estimate-quality assessment. In safety-critical navigation it must support a further decision: whether the information can continue to be used under the specified operational risk and warning requirements \cite{esaNavigationRequirements,wgc2015milestone2}. That decision problem is the subject of navigation integrity in Section~\ref{sec:integrity}.

\section{Navigation Integrity}
\label{sec:integrity}

\subsection{Integrity as an Operational Risk Requirement}

Section~\ref{sec:error-reports} showed why an operational decision cannot be based on the unobservable realized error alone. Integrity addresses that decision for a specified navigation operation. 

Let the operation have an alert limit $AL>0$, an allowed time-to-alert $TTA$, and a prescribed exposure interval. The alert limit is the error tolerance beyond which the navigation information should not remain available for the intended task; $TTA$ is the maximum permitted delay from the onset of the out-of-tolerance condition to an effective warning~\cite{esaNavigationRequirements,wgc2015milestone2}. Let $\mathsf{H}_{\mathrm{op}}$ denote the event that, during the specified operation, the navigation information is declared available, the actual error is beyond $AL$, and the user is not effectively warned to stop using it within $TTA$. For the operation and warning conditions specified here, the integrity risk is the probability of that event~\cite{wgc2015milestone2, milner2020time}:
\begin{equation}
\label{eq:operation-risk}
p_{\mathrm{HMI}}:=\Pr(\mathsf{H}_{\mathrm{op}})\leq IR_{\mathrm{req}},
\end{equation}
Here HMI denotes hazardous misleading information, and $IR_{\mathrm{req}}$ is the allowed risk expressed for the same operation unit.

\subsection{Assumptions Underlying an Integrity Argument}

In integrity monitoring, the actual risk $p_{\mathrm{HMI}}$ is not directly observable. An integrity argument uses probability models with assumptions to evaluate or bound it~\cite{wgc2015milestone2}.
The representative assumptions considered here are (i) nominal error models, (ii) dependence and correlation between error sources, (iii) fault hypotheses, and (iv) fault probabilities~\cite{wgc2015milestone2,blanch2015baseline,milner2020time}.
Together they specify how normal errors, shared error structure, relevant abnormal modes, and their occurrence likelihoods enter the argument.

A model-based risk estimate derived from these assumptions is not automatically the actual operational risk. Its relation to the operation becomes meaningful only when the nominal error models, dependence and correlation, fault hypotheses, and fault probabilities represent the relevant operation. For example, conservative Gaussian overbounds for heavy-tailed GNSS errors can inflate protection levels and reduce availability~\cite{yan2026jackknife}. Temporal exposure also matters: single-epoch and operation-level risks need aligned event definitions and accounting for dependence across time~\cite{milner2020time}.

Thus, the integrity requirement answers how small the hazardous unwarned-use risk must be, while the integrity argument supplies a model-based route for relating observable information to that requirement. The remaining question is what evidence supports the error, uncertainty, and fault characterizations on which that route depends. Section~\ref{sec:credibility} addresses this question through estimator credibility.

\section{Credibility}
\label{sec:credibility}

\subsection{Credibility of Uncertainty and Risk Characterizations}

Section~\ref{sec:integrity} identified the nominal error models, dependence structure, fault hypotheses, and fault probabilities that an integrity argument must use. These assumptions generate reports about positioning uncertainty or navigation risk. In model and simulation assessment, credibility judgments draw on validation, input-data pedigree, uncertainty characterization, and the processes used to produce the results~\cite{babula2009credibility}. Here, credibility asks what justifies such a report: whether the evidence usage and modeling process that produced it adequately support the statistical claim being made.

The claim may take several forms. A dispersion report can state a covariance or central second moment, for example
\begin{equation}
\Sigma=\mathbb{E}[(e_t-\mu)(e_t-\mu)^{\mathsf T}],
\end{equation}
where $\mu=\mathbb{E}[e_t]$. A coverage report can state that a region $C_{0.95}$ contains the error with its nominal probability, such as $\Pr(e_t\in C_{0.95})\geq0.95$. A risk report can state a conservative bound, such as $p_{\mathrm{HMI}}\leq IR_{\mathrm{req}}$, for the specified operation. These claims describe dispersion, coverage, and rare-event risk, respectively, but each is a statistical characterization of the actual error or operational process.

In the estimator literature, credibility concerns whether self-assessed error information is reliable relative to realized estimation performance~\cite{li2002credibility}.
In this broader navigation scope, the same question extends to uncertainty and risk characterizations used within an integrity argument.

\subsection{Sources of Credibility Failure}

One source of credibility failure is evidence usage: how available information is selected, combined, and weighted to support the report. Unaccounted dependence between error sources can lead to understated combined uncertainty. The assessment should also examine whether stale or selective evidence omits relevant operating conditions and whether inappropriate weighting gives a narrow subset of observations undue influence. These problems change the effective support of the evidence-to-claim inference even when the calculation applied to that evidence is internally correct.

A second source is model adequacy: whether the model used to generate the report represents the error-generating process with sufficient fidelity. NMM concerns a reported dispersion that understates or overstates the actual dispersion. SMM concerns a mismatch in the assumed system structure, which can produce a location shift, bias, or other systematic deviation in the estimation error~\cite{yan2026credibleuq}. A credible assessment must keep these mechanisms distinct because a dispersion mismatch and a structural model error call for different evidence and corrective actions.

\subsection{Representative Credibility Diagnostics}

When reliable reference truth is available, normalized estimation error squared (NEES)~\cite{chen2024autotuning} and noncredibility index (NCI)~\cite{li2002credibility} are commonly used to compare realized estimation errors with the estimator's reported second-moment information. When the true state is unavailable, innovation or normalized innovation squared (NIS) tests can be used instead to compare measurement innovations with the assumed innovation model; these provide online consistency tests~\cite{piche2016consistency}. In parallel, the probabilistic forecasting and machine learning communities assess probabilistic predictions using proper scoring rules, such as negative log-likelihood (NLL) and the energy score (ES)~\cite{gneiting2007strictly}.
These scores jointly assess calibration (the statistical consistency between predictive distributions and observed outcomes) and sharpness (the concentration of the predictive distributions), providing complementary evidence for the credibility of uncertainty reports~\cite{gneiting2007calibration}.

\section{Integrity and Credibility: Distinction and Relationship}
\label{sec:comparison}
Sections~\ref{sec:integrity} and~\ref{sec:credibility} clearly demonstrate that integrity and credibility answer different questions. Integrity asks whether hazardous, unwarned navigation use is sufficiently controlled for a specified operation; credibility asks whether the reported uncertainty or risk characterization and its justification are adequately supported. The relationship can be written as evidence and modeling supporting a statistical characterization, that characterization entering an integrity argument, and the argument being judged against an operational risk requirement. A characterization may be credible while showing that uncertainty is too large or risk is too high for continued use, so credibility alone does not establish integrity. Conversely, an integrity claim may be the operational conclusion of an argument whose error and fault assumptions still require a credibility assessment. Credibility therefore concerns the justification of the claims used in an integrity argument, whereas integrity concerns whether the resulting operational risk satisfies the requirement of the specified navigation operation.

\begin{table}[htbp]
\centering
\caption{Comparison of credibility and integrity.}
\label{tab:comparison}
\begin{tabularx}{\linewidth}{@{}>{\raggedright\arraybackslash}p{0.17\linewidth}*{3}{>{\raggedright\arraybackslash}X}@{}}
\toprule
Assessment & Object and question & Probability or evidence & Operational role \\
\midrule
Credibility & Is the uncertainty or risk report justified? & Reported claim, evidence usage, and model adequacy & Supports interpretation of the report \\
Integrity & Is hazardous unwarned use controlled? & Operation-level risk, availability, and warning timing & Assesses compliance with the operational risk requirement \\
\bottomrule
\end{tabularx}
\end{table}
To illustrate the distinction, we list three hypothetical scenarios below: 

\begin{enumerate}
    \item Positioning error may increase while the system enlarges its uncertainty report on the basis of evidence, and a protection-level decision may withdraw the information with an effective warning within the prescribed deadline; realized accuracy is then poor, but the uncertainty characterization can remain credible and hazardous unwarned use is avoided.
    \item The current error may be small while an omitted dependence between sensors makes the reported covariance systematically too small; independent validation can contradict that report even though no hazardous event is observed at that instant.
    \item A system may remain declared available while an error exceeds $AL$ for the prescribed duration and no effective warning arrives within $TTA$, so $\mathsf{H}_{\mathrm{op}}$ occurs. That event may reveal unsupported dependence, an omitted fault hypothesis, or an unjustified fault probability, but it may also be a realization of residual risk represented by a supported model; the event calls for investigation but alone does not establish that the population risk exceeds $IR_{\mathrm{req}}$.
\end{enumerate} 

These scenarios show that realized positioning performance, the credibility of an uncertainty or risk characterization, and operational integrity are related but distinct dimensions of navigation evaluation.

\section{Conclusion}
\label{sec:conclusion}

In safety-critical navigation, assessment cannot rely only on a position estimate or an observed accuracy value; it must also characterize the uncertainty and risk information used to support operational decisions. Integrity concerns whether hazardous, unwarned navigation use remains within the risk requirement of a specified operation, while credibility concerns whether the uncertainty or risk claims used in that assessment are justified by their evidence usage and modeling process. Credibility thus supports the claims on which an integrity argument relies, and integrity determines whether the resulting operational risk is acceptable for the intended use. A complete navigation assessment should keep these layers explicit and ask what actually happened, why the reported uncertainty or risk should be believed, and whether that information is sufficient for continued use in the specified operation.

\begingroup
\small
\setlength{\bibsep}{4pt}

\endgroup

\end{document}